\documentclass[fleqn,usenatbib]{mnras}

\usepackage{newtxtext,newtxmath}
\usepackage[T1]{fontenc}

\DeclareRobustCommand{\VAN}[3]{#2}
\let\VANthebibliography\thebibliography
\def\thebibliography{\DeclareRobustCommand{\VAN}[3]{##3}\VANthebibliography}

\usepackage{graphicx, float}	% Including figure files % For figures placement
\usepackage{amsmath}	% Advanced maths commands
\usepackage{enumerate}  %To add some enumerating styles
\usepackage{multirow}   % Work with tables      
\usepackage{array}      % Work with arrays
\usepackage{caption}    % To write footnotes to images
\usepackage{subcaption}
\graphicspath{{./figures/}} % Path to figures
\usepackage[dvipsnames,table]{xcolor}
\usepackage[normalem]{ulem}
\newcommand{\reffig}[1]{Fig.~\ref{fig:#1}}

\newcommand{\spine}{$\texttt{SpineWeb}$ }
\newcommand{\mpch}{$h^{-1}\mathrm{Mpc}$}
\newcommand{\Stop}{{\sc Smooth 1}}
\newcommand{\Smid}{{\sc Smooth 2}}
\newcommand{\Sbot}{{\sc Smooth 4}}
\newcommand{\sage}{{\sc sage}}
\newcommand{\sinv}{\textsc{spine-in-voids}}	
\newcommand{\MthUnit}{$M_\mathrm{th} \sim 10^{12} M_{\odot}h^{-1}$}
\def\simlt{\stackrel{<}{{}_\sim}}

\newcommand{\Modoth}{$M_{\odot}h^{-1}$}
\providecommand\apj{Astrophys.J.}
\providecommand\apjl{Astrophys.J.Lett.}
\providecommand\mnras{M.N.R.A.S.}
\providecommand\aap{Astron.\&Astroph.}
\providecommand\jcap{Journal of Cosmology and Astroparticle Physics}
\providecommand\prd{Physical Review D}

\title[Galaxies tracing the cosmic web]{Hierarchical structure of the cosmic web and galaxy properties}

\author[M. Jaber et al.]{Mariana Jaber$^{1}$\thanks{E-mail: jaber@cft.edu.pl (MJ)}, %\thanks{ORCID:https://orcid.org/0000-0001-7507-9516}
Marius Peper$^{2}$, \thanks{E-mail: mpeper@astro.uni.torun.pl (MP)},
Wojciech A. Hellwing$^{1}$, %\thanks{E-mail: hellwing@cft.edu.pl (WH)}
Miguel A. Arag\'on-Calvo$^{3}$, %\thanks{E-mail: maragon@astro.unam.mx (MAAC)},
  \newauthor and Octavio Valenzuela$^{4}$ %\thanks{E-mail: octavio@astro.unam.mx (OV)}
\\
$^{1}$Center for Theoretical Physics, Polish Academy of Sciences, Al. Lotnik\'ow 32/46, 02-668 Warsaw, Poland\\
$^{2}$Institute for Astronomy, Faculty of Physics, Astronomy and Informatics,  Nicolaus Copernicus University, Grudziądzka 5, 87-100 \\ \-\ Toru\'n, Poland. \\
$^{3}$Instituto de Astronom\'ia, Universidad Nacional Aut\'onoma de M\'exico, Apdo. Postal 106, Ensenada 22800, B.C., M\'exico.\\ 
$^{4}$Instituto de Astronom\'ia, Universidad Nacional Aut\'onoma de M\'exico, A. P. 70-264, 04510, M\'exico, CDMX, M\'exico.\\
}

\date{Accepted 2023 October 23. Received 2023 October 23; in original form 2023 April 27}

\pubyear{2023}

\begin{document}
\label{firstpage}
\pagerange{\pageref{firstpage}--\pageref{lastpage}}
\maketitle

\begin{abstract}
Voids possess a very complex internal structure and dynamics. Using $N$-body simulations we study the hierarchical nature of sub-structures present in the cosmic web (CW). We use the \spine method which provides a complete characterization of the CW into its primary constituents: voids, walls, filaments, and nodes.
We aim to characterize the inner compositions of voids by detecting their internal cosmic-web structure and explore the impact of this on the properties of void galaxies. 
Using a semi-analytical galaxy evolution model we explore the impact of the CW on several galaxies' properties.
We find the number fraction of haloes living in various CW components to be a function of their mass, with the majority of the haloes of mass below $10^{12}$\Modoth, residing in voids and haloes of higher masses distributed mostly in walls.  
Similarly, in the Stellar-to-Halo mass relationship, we observe an environmental dependence for haloes of masses below  $10^{12}$\Modoth, showing an increased stellar mass fraction for the galaxies identified in the densest environments.
The spin is lower for galaxies in the densest environments for the mass range of  $10^{10}-10^{12}$\Modoth.  Finally, we found a strong trend of higher metallicity fractions for filaments and node galaxies, with respect to the full sample, in the range of  $M_*\simlt10^{10}$\Modoth. 
Our results show that cosmic voids possess an intricate internal network of substructures. 
This in turn makes them a complex environment for galaxy formation, impacting in an unique way the properties and evolution of the chosen few galaxies that form inside them.
\end{abstract}

\begin{keywords}
cosmology: large-scale-structure -- methods: numerical -- galaxies: evolution %cosmic-web -- galaxy properties
\end{keywords}

%%%%%%%%%%%%%%%%%%%%%%%%%%%%%%%%%%%%%%%%%%%%%%%%%%

% ========================================================= 
% ================== INTRODUCTION  ========================
% =========================================================
\section{Introduction}

Cosmic voids are one of the most noteworthy elements of the distribution of matter in the Universe on mega-parsec scales. Along with walls, filaments and nodes, they constitute an interconnected network of the cosmic web. 
The theory describing the formation and evolution of the cosmic web is rooted in the Zel’dovich formalism \citep{zeldovich70, doroshkevich_shandarin_78}. The first detection of this profound large-scale arrangement of matter in the Universe can be traced back to early galaxy surveys and the works of \cite{chincarini_rood_75, gregory_thompson_78, kirshner_1981, kirshner_87}. 
From the compilation of the early catalogues \citep{2003astro.ph..6581C, 2000AJ....120.1579Y, 2009ApJS..182..543A} to present day, the study of cosmic voids has consolidated their potential as a powerful cosmological probe.

It has been demonstrated that the study of the properties and distributions of cosmic voids offers great prospects for cosmological model and theory testing. 
Among many applications, a few successful examples would include: 
the void-size function as a tool to constrain cosmological parameters \citep{sheth_RvdW_2004, jennings_void_excursion, pisani_void_counts, contarini_voidsizefunc_bias}, 
the measurement of cosmic shear around voids in weak-lensing studies \citep{krause_wl_voids,shed_voids1, shed_voids2, des_y1_wl}, 
the cross-correlation of stacked voids with temperature maps from the CMB \citep{2008ApJ...683L..99G,2016ApJ...830L..19N,des_y1_voids}, 
testing the Alcock-Paczy\'nski effect \citep{ap_voids, ap_voids_sdss}, studying the cross-correlation of void centers with positions of luminous tracers around them, serving as a probe of growth function in redshift-space distortions (RSD) analysis \citep{cai16,sesh_will, sesh_prd19, sesh_19,  sesh_voids_lrg, correa_rsd_voids1, sdss_voids}, and studying the fractal dimension of the cosmic web \citep{topology_cweb_rien, persistent_betti_cweb}, or more recently, the scaling relations based on their connectivity \citep{Aragon23-voids}.
Furthermore, understanding the processes that produce and shape the cosmic web, can be key in understanding the validity regime of gravity as prescribed by the theory of General Relativity \citep{shed_voids1, shed_voids2}.

The cosmic voids are, by the nature of gravitational collapse, large-scale structure phenomena. Because of this, they have been mostly and primarily used as tools to study the background cosmology, owing to their sensitivity to the underlying background model and the large-scales. However, a different perspective arises from the realisation that cosmic voids comprise one of the most extreme environments for galaxy formation. This might appear as a bit surprising view, considering that voids sometimes are defined as regions deprived of galaxies. Indeed, the cosmic voids are almost empty regions, but they are not totally empty (see for instance \cite{pan2012}). The standard cosmological model predicts that there should be a meager population of small galaxies that form and evolve inside the voids.
There have been several studies investigating
the effect of a unique void environment on their properties \citep[see for e.g.][]{2002ApJHoyle.Vogeley,2004ApJ...617...50R,Patiri2006, Kreckel2011, 2016_Beygu, 2021_Florez, 2022_DominguezGomez}, and references discussed in \citep{sheth_RvdW_2004}.
It has also been shown that this unique cosmic environment has significant impact on the formation and evolution of dark matter haloes and their intrinsic properties. 
General consensus being that formation and virialisation of haloes in voids is delayed, which is reflected by lower central densities, lower total angular momenta and less spherical density profiles \citep{vladimir, 2017MNRAS.469..594B, hellwing21}.

Some other works focus on analysing the internal structure of cosmic voids, either using high-resolution $N$-body simulations \citep{Gottloeber2003, Aragon_Calvo_2012}, but also detecting it directly in the observed distribution of galaxies \citep{1991MNRAS.248..313K, 2004ApJ...617...50R, 2006ApJ...640L.111T, gama_filaments_in_voids, 2017ApJ...835..161M}. 
While several studies have investigated the internal structure of cosmic voids and the impact of galaxy placement in the cosmic web, separately, there remains a need for a comprehensive analysis of these factors together. More precisely, while works such as \cite{2010MNRAS.404L..89A,aragon_calvo_multiphen,Aragon_Calvo_2012}    looked at the internal composition of voids using similar methods as us, and other papers employed different techniques to look at the internal structure of cosmic voids and their impact on DM haloes  \citep{Gottloeber2003, Cautun_Nexus, hellwing21}  the novelty of our work lies in lining the hierarchical multi-scale description of the cosmic web, with their effect on the galaxy properties, going beyond the mere analysis of DM haloes or sub-haloes. 

The goal of this paper is to address this gap and provide insights into the internal structure of voids. We present a novel analysis of the interconnection between galaxy properties and the internal properties of voids in cosmological simulations. First, we focus on the internal structure of voids identified in gravity-only simulations, and the cosmic web segmentation into voids, walls, filaments and nodes. Then, we move to analyze the physical properties of the haloes and galaxies formed in each cosmic web component: size, spin, stellar mass, and metals content. Specifically, we aim to determine which type of galaxies traces the internal filamentary  (and associated voids, walls, and nodes) structure of voids, extending the work of \cite{Aragon_Calvo_2012} by looking at galaxy properties in the context of the hierarchical structure of cosmic voids. By addressing this question, we seek to deepen our understanding of the processes that shape the largest structures of the Universe and the formation and evolution of galaxies residing in each component.

The structure of the paper is as follows. In section \ref{sec:methods} we describe our set of $N$-body simulations, and the algorithms used to interpolate the density field as well as to segment it into the cosmic components. We go over our methodology to populate dark-matter haloes with galaxies in \ref{subsec:sage}. 
Section \ref{sec:results} contains our main results concerning the properties and inner structure of voids, which we discuss in section \ref{sec:discussion}. In Sec. \ref{sec:conclusions} we present our conclusions.

%-----------FIGURE 1: dens fields ------------------------
\begin{figure*}
\centering
	\includegraphics[width=\textwidth]{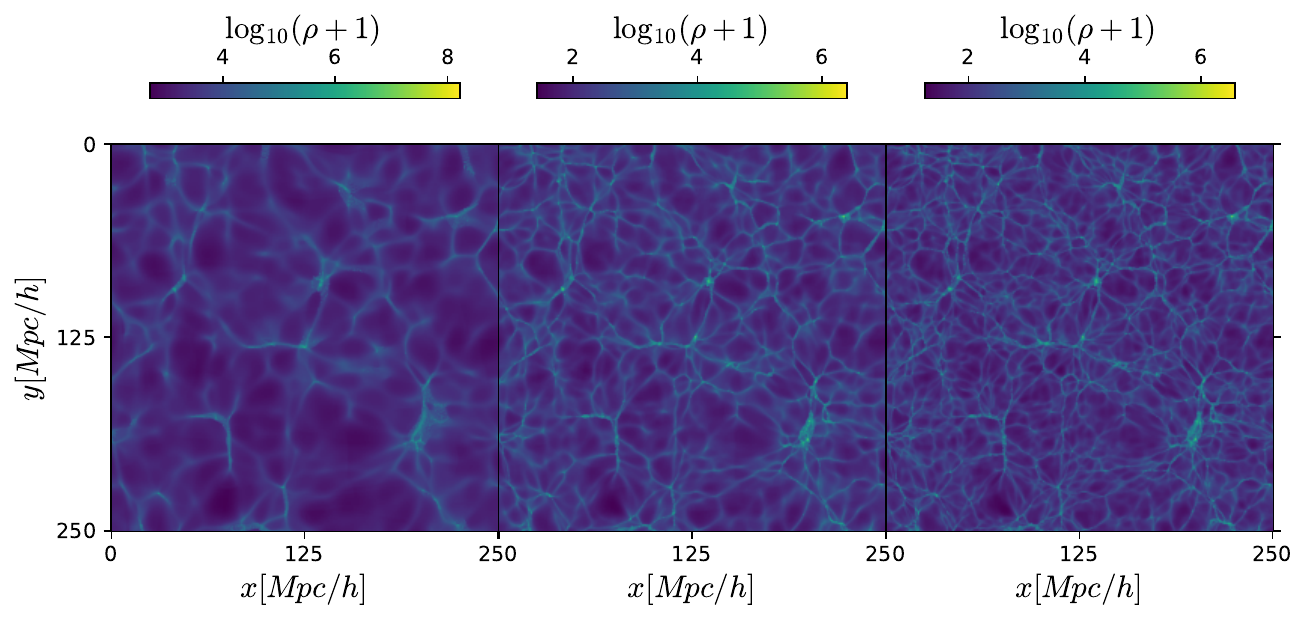}
    \caption{A thin slice of the estimated density field from the simulation boxes at $z=0$ for the three hierarchical levels: from left to right, \Sbot{}, \Smid{}, and \Stop{}, \-\ respectively. 
    The effect of the designed filters at $\beta_s = {4, 2, 1}$\mpch \-\ is noticeable as we move from left to right.}
    \label{fig:density124}
\end{figure*}
%---------------------------------------------

\section{Methods}
\label{sec:methods}

The cosmic voids are phenomena of the large-scale matter density distribution in the Universe. Since the density is a scale-dependent quantity, the definition and the connected void-finding methodology is a subject of noticeable variation in
the field (see eg. \cite{aspen_amsterdam, libeskind_comparison, shed_voids1}). 
For the sake of reproducibility and admissibility of comparisons of the results, it is important to clearly define and describe our methodology. This section is devoted to this goal.

% ========================================================= 
% ===================== s1 s2 s4  ========================= 
% =========================================================
\subsection{\label{subsec:hierarchy} Hierarchical space construction}
Some studies identified voids directly on the galaxy distribution in order to match observations.
However,  since our focus lies on describing the inner structure of voids, using the dark-matter (DM) particle distribution rather than the galaxy distribution is better suited for this analysis.

The analysis presented here is formally based on the Bolshoi simulation\footnote{\url{https://www.cosmosim.org/metadata/bolshoi/}} with $ h=0.7 $, $\sigma_8 = 0.82$, and box size, $L= 250$\mpch{} and a mass resolution of $8.64\times10^9$\Modoth (the original Bolshoi run has a mass resolution of $1.35\times10^8$\Modoth, \cite{2011ApJ...740..102K}). However we use the Bolshoi initial conditions re-sampled on a regular $512^3$ particle grid. 
In order to be able to identify voids at different scales and reconstruct their hierarchical relations we construct a hierarchical space which is a generalization of Gaussian smoothing-based scale-spaces \citep{aragon_calvo_multiscalemorphfilter, Cautun_Nexus} but explicitly containing hierarchical relations while retaining the anisotropic features of the cosmic web structures. 

A hierarchical space is constructed by removing small-scales from the initial conditions of the simulation and then running this smoothed initial conditions as a regular simulation. We use a \emph{sharp}$-k$ filter as this allows us to cleanly target specific scales to be removed. 
This filter does not introduce artifacts as it is applied in the linear regime where the Fourier modes are independent 
\citep{einasto_wavelet,sheth_RvdW_2004, Sheth_2004_1, einasto11}. 

Applying such filter one can remove the small-scale details and evolve the filtered density-field to obtain the full non-linear array of substructures above the cut-off scale, $\beta_s$.

The resolution in our simulations limits us to three cut-off scales, $\beta_s = {4, 2, 1}$\mpch, which we applied to each one of the simulation boxes and the resulting particle distributions are labeled \Sbot{}, \Smid{}, and \Stop{}, respectively (see Fig. \ref{fig:density124}). 
%

% ========================================================= 
% ========================= LTFE  ========================= 
% =========================================================
\subsection{\label{subsec:ltfe} Density field reconstruction}

From the final particle distribution at $z=0$ we estimate the corresponding density fields on a cubic grid of resolution $N_g = 512^3$. Additional steps in order to ensure good quality density fields and to remove small-scale spurious variations or any type of artifacts can be taken: voxels (cubic pixel) with density values ten times higher than the median of their adjacent voxels are substituted by the mean and the density field is smoothed with a Gaussian filter of $\sigma = 2$ voxels. The value of these kernels are the only free parameters in the \spine{} framework.

As our motivation lies in studying the internal structure of voids we require a method that excels in reconstructing the density field in the under-dense regions.
For this reason we have used the CPU implementation\footnote{ \url{https://github.com/miguel-aragon/Lagrangian-Sheet-Density-Estimator}} of the Lagrangian Sheet Density Estimator (LSDE) method, originally proposed in \cite{Shandarin_2011, Abel2012}. The work of \cite{wojtak2016} follows the same procedure for estimating their density fields. 

To implement the method, the first step involves decomposing the particle grid into tetrahedra in order to compute slices across the 3D Lagrangian tessellation taking advantage that tetrahedra can be used to approximate complex three-dimensional shapes, as they can tessellate space efficiently. This is followed by rendering the tetrahedral decomposition into a data-cube using a triangle-renderer. For each voxel there are eight possible tetrahedral partitions, which are all computed and then the density at each vertex is the resulting average.  We chose a cubic grid of 512 cells per size, giving voxels of side-length of 488$h^{-1}$kpc.

%%%%%% =========================

\subsection{\label{subsec:cweb}Cosmic web segmentation}

The cosmic web is made up of vast, interconnected networks of galaxies and dark matter, which are arranged in a web-like structure that spans the entire observable Universe. This structure has a distinct topology, meaning that it has a specific pattern of connectivity, shape, and arrangement.

In order to understand the cosmic web and its properties, it is important to use methods that take into account its topological nature. 

The method we use for identifying the cosmic web components is the \spine{} algorithm  \citep{spineweb} which has been specifically designed to capture and analyze the complex, interconnected structure of the cosmic web and takes into account its topology in a comprehensive and accurate way. The \spine{} algorithm is based on the watershed segmentation of the density field \citep{platen_2007}. 
\spine{} as a topological cosmic web segmentation method is an application of the Morse Theory. \cite{libeskind_comparison} compares a number of  different cosmic-web segmentation approaches, including topological algorithms such as \spine{} and \textsc{Disperse} \citep{2011MNRAS.414..384S}.  These kind of methods study the location and type of critical points of the density field, and recover useful information about the number of connected components, their boundaries, and the topology of the cosmic web. While \spine{} excels for the identification of void regions, \textsc{Disperse} performs exceptionally well for the detection and study of cosmic filaments. A brief compilation of filament finders and a new algorithm is presented in \cite{2022MNRAS.514..470P}.

\spine{} begins by identifying all the voxels  with the lowest density value among all their 26 neighbours, which will be flagged as the seed for the cosmic voids. 
Then it runs a topographical distance algorithm finding the maximum gradient paths.
After all the voxels are connected to the lowest-density neighbours, the standard flooding algorithm marks the points where two or more watershed regions converge.  
By identifying all the voxels lying on the watershed boundaries, a morphological label can be attached to discriminate between voxels belonging to a void, a wall,  a filament, or a node.
This is done by taking advantage of the fact that every voxel in the segmentation is assigned an index numbering the different watershed regions, so by counting the number of neighbouring voids,  $N_{v}$, adjacent to a watershed region, we classify each voxel in the boundary as belonging to a void. 

This morphological label is assigned by counting the number, $N_{v}$, of voids neighbouring the watershed region.
The criteria we use to assign this label are  the following: if there are no other voids than the current one (or $N_{v}=1$), it means we are inside of a void;  $N_{v} = 2$ or having two different adjacent voids means that we are lying in a wall; having $N_{v} = 3$ different voids is classified as a filament, and having $N_{v}  \geq 4 $, is classified as a node. As every voxel is assigned to a watershed region, the case $N_{v}=0$ or no voids found, is not possible. 

As a result we have our density field segmented into voxels accordingly labelled as voids, walls, filaments and nodes, which we can merge to obtain the full description of the cosmic web among all three hierarchical scales. 

As a final step, after the segmentation into individual cosmic web elements, \spine{} constructs a hierarchy of voids identifying the parent voids (the largest ones, found at the bottom of the hierarchy or  \Sbot{}, as seen in the right panel of \reffig{density124}), and merging them with the structures identified in the intermediate hierarchical level, \Smid{} (middle panel of \reffig{density124}), and finally, with the structures present in the top of the hierarchy or \Stop{} (left panel of \reffig{density124}). 
This step ensures that the voids identified in the bottom of the hierarchy have a watertight connection in terms of the smaller voids and cosmic web elements identified in the middle and the top of the hierarchy. It is thanks to this connection that we can describe the internal composition and structure of cosmic voids identified at \Sbot{} in terms of the voids, walls, and filaments identified at \Stop{}. For the full details of this technique, we refer the reader to \citep{spineweb,Aragon_Calvo_2012}. 

%-----------FIGURE 2: void sizes ------------------------
\begin{figure}
	\includegraphics[width=\columnwidth]{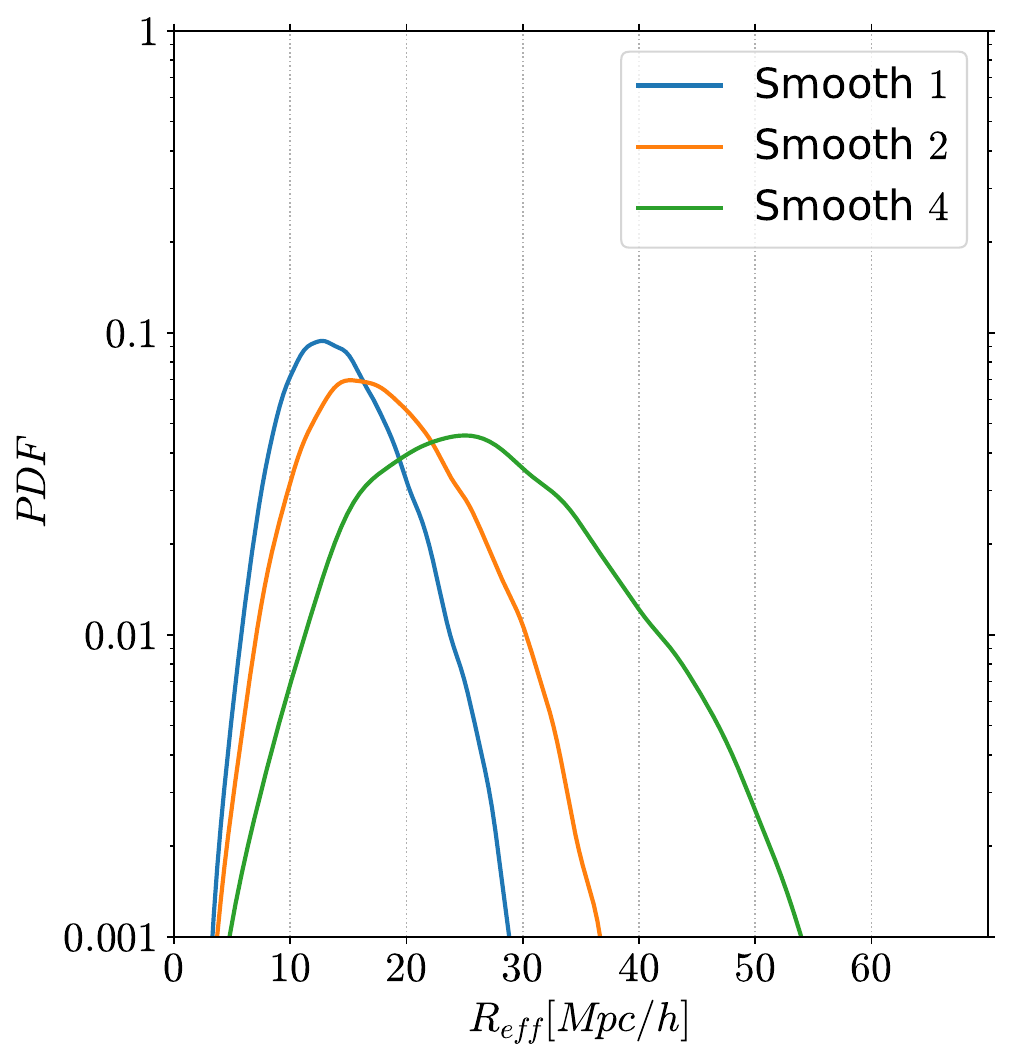}
    \caption{Probability distribution functions for the void sizes in the three hierarchical scales.}
    \label{fig:pdfs}
\end{figure}
%---------------------------------------------

%-------------FIGURE 3: dens profiles  ---------------------

\begin{figure*}
    \centering
    \begin{subfigure}[b]{0.32\textwidth}
        \includegraphics[width=\textwidth]{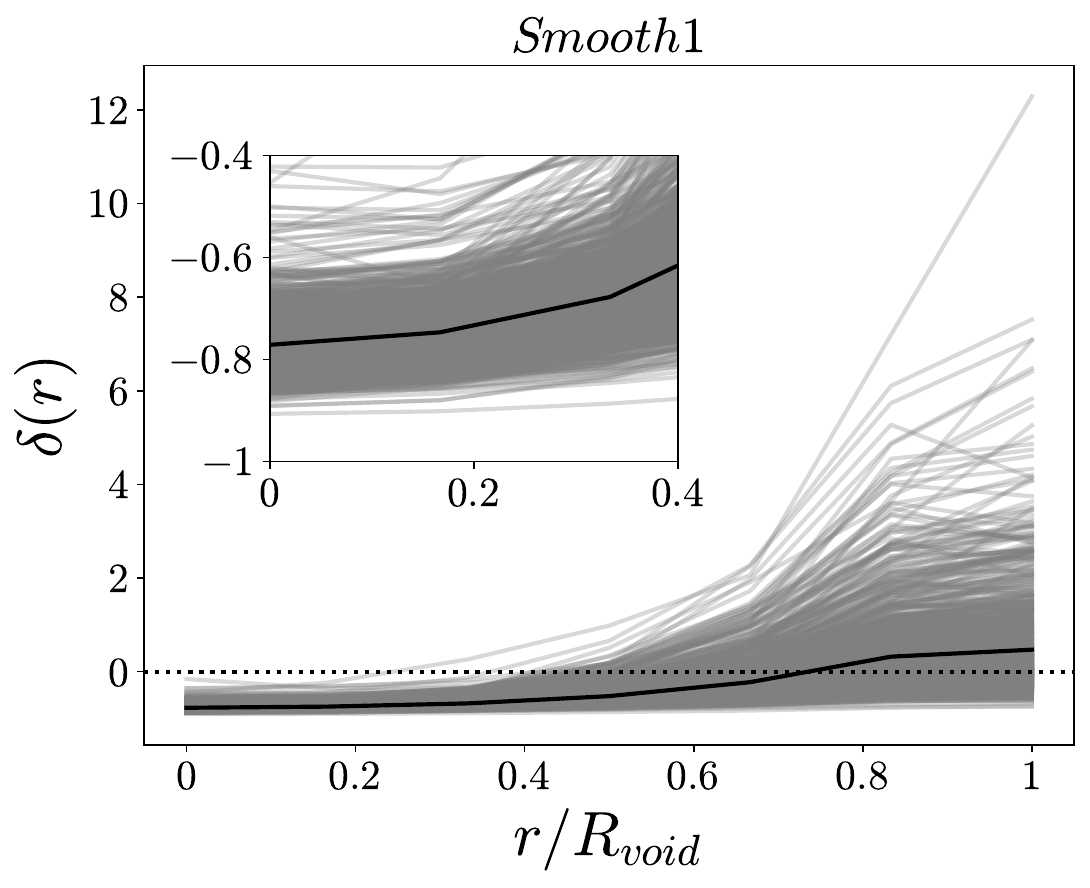}
        \label{fig:dens_s1}
    \end{subfigure}
    \begin{subfigure}[b]{0.32\textwidth}
        \includegraphics[width=\textwidth]{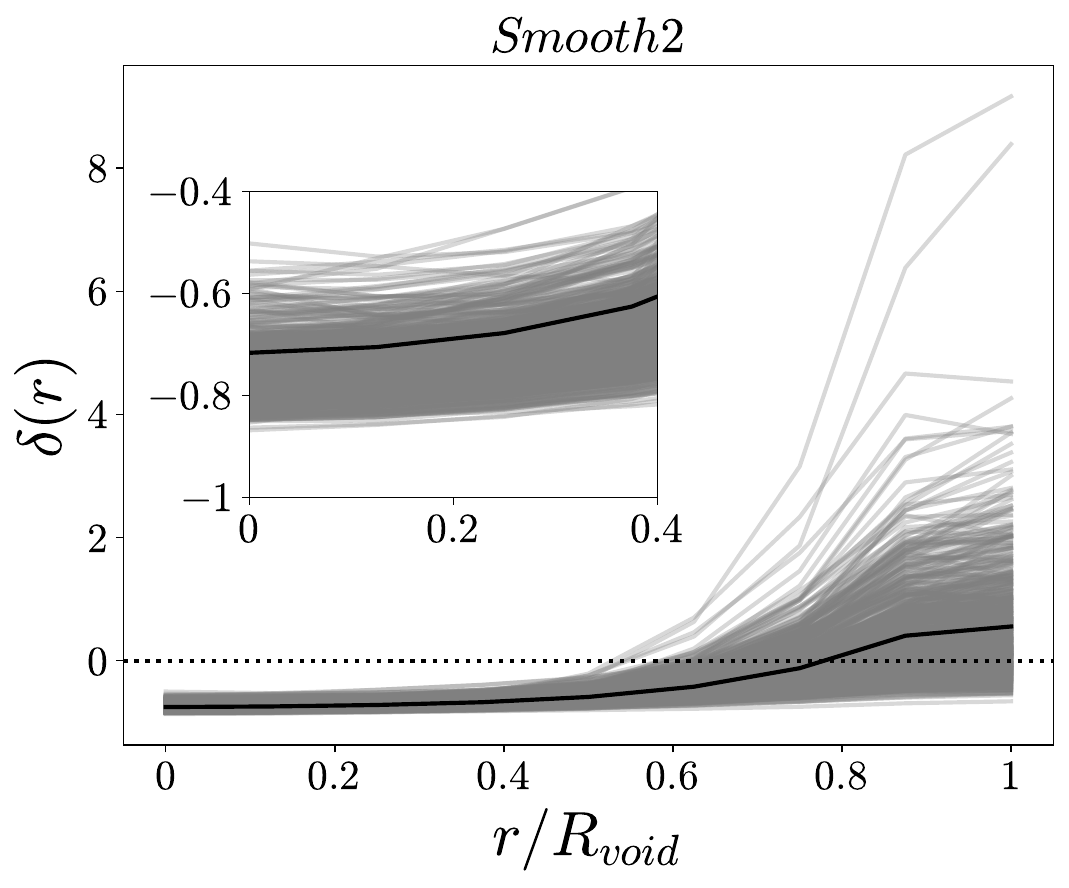}
        \label{fig:dens_s2}
    \end{subfigure}
    \begin{subfigure}[b]{0.32
    \textwidth}
        \includegraphics[width=\textwidth]{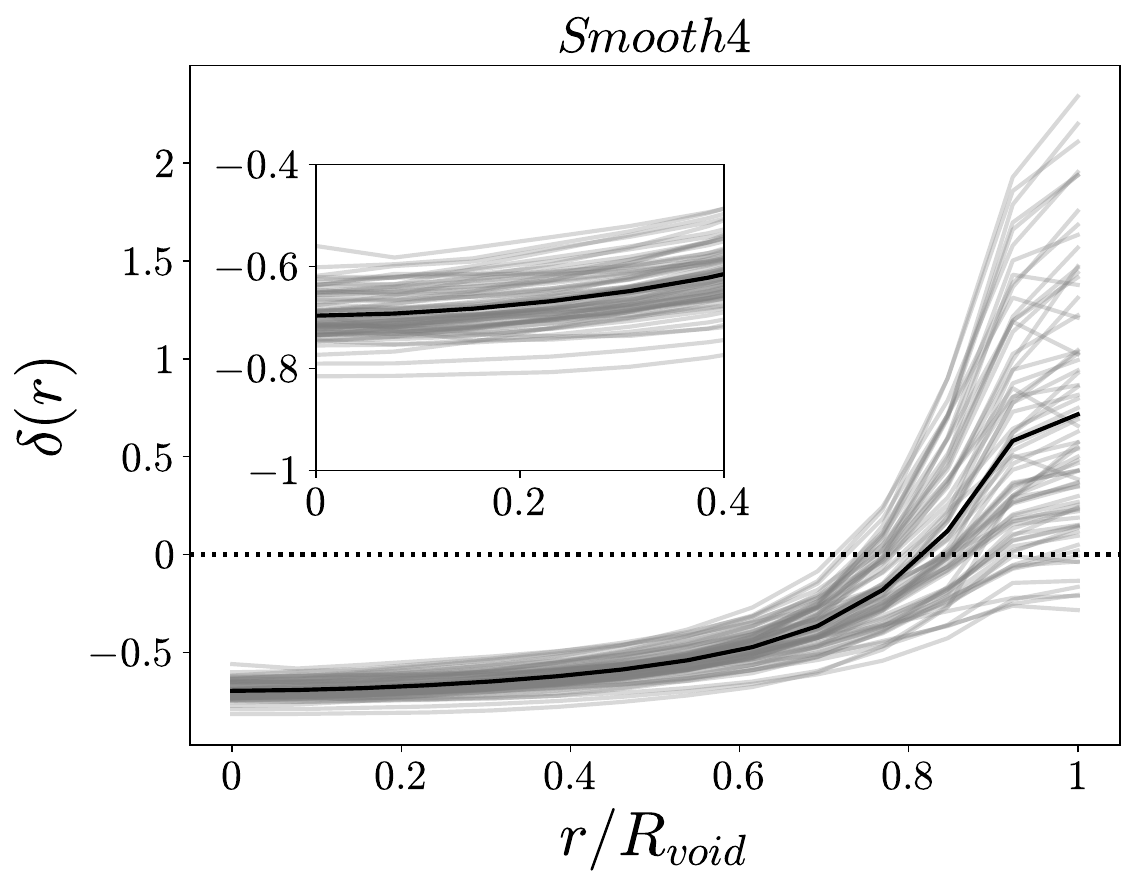}
        \label{fig:dens_s4}
    \end{subfigure}
    \caption{Density profiles for voids across the hierarchy, as function of $R=r/R_\mathrm{void}$. The light-gray lines show several individual void profiles while the black line corresponds to the mean of all the profiles. For each hierarchical scale we selected voids with $R_\mathrm{void}$ according to the peak of the distributions in \ref{fig:pdfs}. The profiles follow the typical bucket-shape form and have the characteristic value of $\delta \sim-0.8$ at the center of the voids, as illustrated in the inset plots. }
    \label{fig:densprofiles}
\end{figure*}
%--------------------------------------------
% ========================================================= 
% ========================= sage  ========================= 
% =========================================================
\subsection{\label{subsec:sage} Semi-Analytical Galaxy Evolution} 

Now, in order to characterize the galaxy population tracing each cosmic environment we used the publicly available semi-analytical galaxy evolution software \sage{} \citep{Croton16SAGE}\footnote{\url{https://github.com/sage-home/sage-model}}.

\sage{} was designed to work on every $N$-body simulation with appropriately formatted merger trees, which contain full information of the hierarchical growth of the dark matter haloes over cosmic time. Semi-analytical tools then populate each halo with galaxies, using empirical relations and physical models that are carefully fine-tuned to match the observations.
\cite{Baugh06review} gives a comprehensive review on semi-analytical tools, and \cite{2018MNRAS.474.5206K} explores the performance of \sage{} and two other semianalytical evolution codes.

To do so, \sage{} models the baryonic processes that are thought to be important for galaxy formation and evolution, such as gas cooling and heating, gas accretion or ejection, star formation, supernova feedback, black hole growth, and AGN feedback while it keeps track of the gas in satellite galaxies and galaxy mergers as well as the build-up of intra-cluster stars.

Using empirical relations and physical models, \sage{} can predict the properties of galaxies within each halo at different redshifts. The semi-analytical approach models the statistical properties of galaxies very well, however, a detailed discussion of the processes in a single galaxy are beyond the scope of our analysis.

For this part of the analysis we took the halo merger-trees for the full Bolshoi data, which were generated using the friends-of-friends phase space halo finder {\sc rockstar} \citep{Behroozi13rockstar} and {\sc consistent-trees} \citep{Behroozi13MHT}. 
For the Bolshoi simulation this data is publicly available\footnote{\url{https://www.slac.stanford.edu/~behroozi/Bolshoi_Trees/}}. The fiducial \sage{} parameters used are described in table 2 of \cite{Croton16SAGE}, tuned specifically for the case of the Bolshoi simulation.

The matching between the three different hierarchical scales is guaranteed by construction (see \ref{subsec:hierarchy} and the extensive description found in \cite{spineweb,Aragon_Calvo_2012}). 
To validate the connection between our galaxy sample and the cosmic web segmentation, we cross-referenced the positions of each galaxy with the density values in the set of three low-resolution Bolshoi analogues. 
Our analysis revealed a strong agreement between high-density regions and the spatial concentration of galaxies, and conversely, low-density regions and a lower concentration of galaxies. 
Therefore we proceed describing the physical properties of galaxies in the context of our cosmic-web segmentation.

The galaxy properties we monitor are: the dimensionless spin parameter, $\lambda$, the masses of host halo, the stellar-to-halo mass relation, and the  galaxy metallicities. This allows for an analysis of the galaxies residing in different environments, namely in voids, filaments, walls and clusters, but also, those tracing the structures found in the inner regions of the larger voids. 
%
% ========================================================= 
% ======================= RESULTS  ========================
% =========================================================

\section{Results}
\label{sec:results}
We present our results related to the hierarchy of the cosmic web in three stages. First we focus on the properties of the reconstructed density fields and the multi-scale nature of the structures imprinted after each filter was applied. Then we discuss their impact on the void properties: void sizes distribution across the hierarchy, and their density profiles. 
Lastly, we focus on the relationship between the physical properties  of haloes and galaxies, and their location in different cosmic web environments.

% =============================================== 
% ============    density fields  ===============
% ===============================================
\subsection{Density fields}

We illustrate the manner in which the hierarchical structure of the cosmic web is revealed using the filters  $\beta_s = {1, 2, 4}$\mpch,  by presenting a thin slice of the density fields in figure \ref{fig:density124}. 
The right-hand panel of the figure shows the density field for the simulated volume applying the least restrictive filter, \Stop{}, where all the large and small-scale details are present in the tenuous filaments and walls surrounding the voids. The middle panel shows the same slice but for the \Smid{}  density field, where only the modes larger than $\beta_s=2$\mpch{} in the initial power spectrum have evolved and only the structures growing from these modes are preserved. 
Finally, the left-hand panel shows the bottom of the hierarchy, or \Sbot{}, where  all the modes below $\beta_s=4$\mpch{} are absent. In other words, we see that the same large-scale features are present in all 3-levels, be it big voids, massive density knots or elongated filaments. However, by moving from left to right (\textit{i.e.} from \Stop{} to \Sbot{}) we can observe gradual emergence of smaller ``features within features". This simple exercise illustrates the multi-scale nature of the cosmic-web, where all large-scale phenomena such as voids, filaments and even density nodes, contain smaller substructures.

% ========================================================= 
% ====================   voids properties   ====================
% =========================================================
\subsection{Void size distribution and density profiles}

It is well established that  the matter distribution in realistic cosmic voids will never reach a spherically-symmetric arrangement. This is because voids do not live in isolation, but rather are a subject of the tidal-fields induced by the local large-scale structures  \citep{2009AAS...21342507P, 2009ApJ...699.1252F,van_de_Weygaert_2014}. 
However, a commonly adopted approach to characterize  void sizes, is to use an effective radius of a sphere containing the same volume, $V$, as a given void. This is more convenient than working with 3D volumes themselves, and since majority of voids have quite regular shapes, provides us also with a useful notion of a given void spatial scale.
The volume $V$ of a void using the watershed method is computed as the sum of the volumes of all the voxels constituting the void.

We show the distribution of these effective void radii for our three levels of hierarchy in figure \ref{fig:pdfs}.
We notice how the three curves are very similar and resemble a log-normal distribution.
For the top of the hierarchy, the peak of the distribution is around $\sim 12$\mpch, and it moves as we move across the hierarchy, to $\sim15$\mpch \-\ and $\sim25$\mpch.  
We find that the peak scale for each level scales with the smoothing in a simple relation as $R_\mathrm{peak}\sim 11\beta_s^{0.6}\textrm{\mpch}$. 

According to the prevailing theory, voids  originate from negative density
fluctuations in the primordial density field. Their under-density causes a repulsive force that leads to their expansion. This expansion causes the matter within the voids to move towards their boundaries and results in a void density profile that can be approximated by the shape of a reverse top-hat. 

The density profiles for our sample, shown in figure \ref{fig:densprofiles}, follow the characteristic bucket-shaped form, and the density contrast at the center  approaches a  value of $\delta\sim-0.8$. 
Thus, our sample of voids reproduces the well established results of \cite{Sheth_2004_1}, where $\delta = (\rho-\bar{\rho})/\bar{\rho}$, with $\bar{\rho}$ the background density of the Universe. 

The void density profiles estimation is done by calculating the average density in a given shell. It starts at the boundary of the void, or the innermost shell, and then moves inwards following the taxicab metric or distance on a regular grid\footnote{We checked that our results do not depend on this choice over the Euclidean distance estimation}.

The innermost region of the void would contain the lowest values of density in the watershed region. On this way we can see that the ``center of the void'' does not require the assumption of spherical symmetry. 
For a discussion about different definitions of void centers, see the Appendix of \cite{sesh_will}.

In the three panels we notice that, as we get closer to the boundary of each void, the density increases several times the value of $\bar{\rho}$. In the right hand panel, this value increases up to twice the background density, in the middle one, up to 8 times, and in the left one, a dozen times the background value.
This is a consequence of two main factors. 
On the one hand, naturally, the intersection between the corresponding sphere of of radius $R_\mathrm{eff}$ and the boundaries of the voids, varies with the void size: for the smaller voids, the sphere with the same volume, $V$, would intersect the void boundaries in couple of points, which are outside of the void  and much more dense. On the contrary, the larger voids tend to be more spherical, and in these cases the deviation from an effective sphere description is weaker. 

However, a more interesting factor is the effect of the smoothing scale. 
We have filtered out the smallest voids in  \Sbot{} (right panel) and \Smid{} (middle panel), while having an abundance of small structures at the top of the hierarchy, or \Stop{} (left panel).  
In the latter case, the resulting voids are smaller but also the  density contrast at their boundaries is larger than in the former cases. In other words, there is a noticeable effect of the web of filaments and nodes delineating the voids, more prominently featured in the unfiltered case (\Stop{}), than in the other two density fields, as we can see in the right-hand ends of the three panels of figure \ref{fig:densprofiles}.

\begin{figure*}
\centering
	\includegraphics[width=\textwidth]{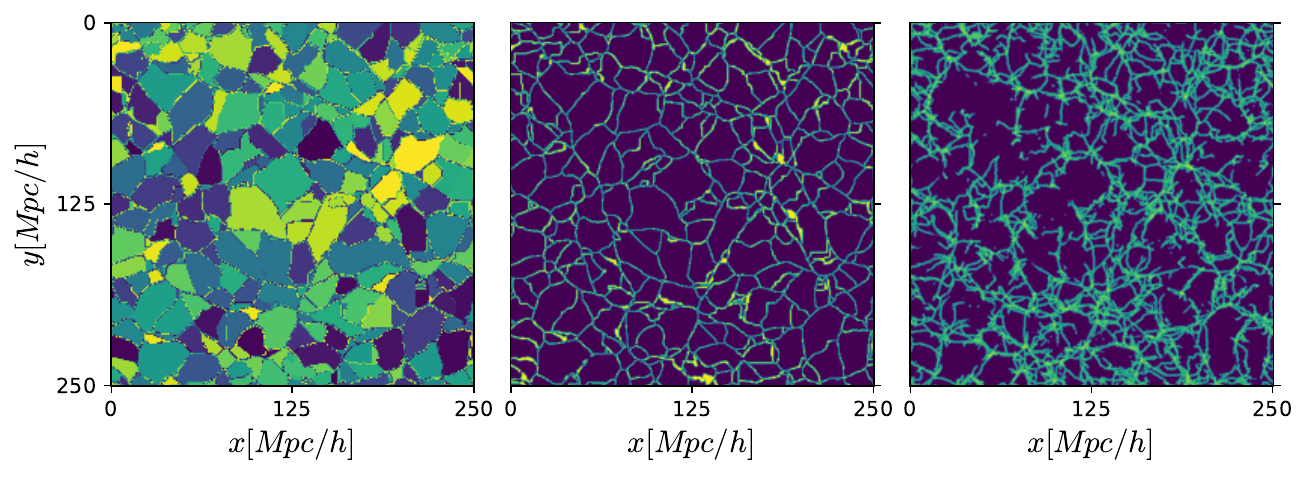}
    \caption{A thin slice of the cosmic web segmentation into voids (left panel), walls (middle panel) and filaments plus nodes (right panel) of the \Sbot{} density field. The tight correspondence of the three environments is evidenced. } 
    \label{fig:spine}
\end{figure*}

The result from the cosmic-web segmentation using the morphological label can be seen in the three panels of figure \ref{fig:spine}. 
The left-hand panel shows the watershed regions identified as voids, while the middle panel clearly shows the voxels identified as walls. The airtight connection between the boundaries of voids and regions identified as walls is evidenced: as no voxel is left unclassified the walls correspond to a perfectly surrounding structure around voids. Finally, the right-hand panel shows the filaments and nodes surrounding the empty regions. 

\begin{figure}
    \centering
    \includegraphics[width=\columnwidth]{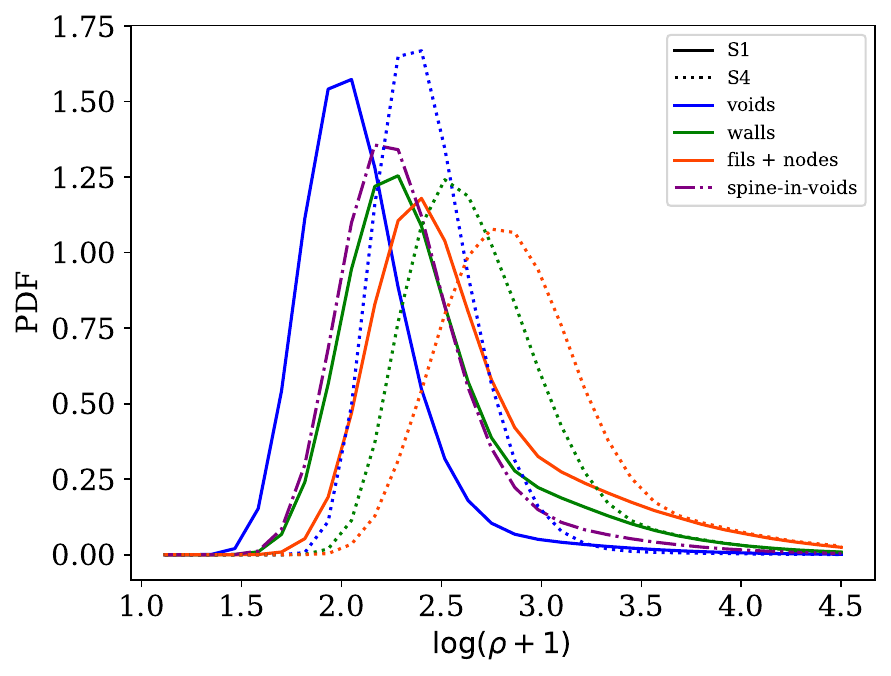}
    \caption{The density PDF in each environment of the cosmic web as detected by by SpineWEB. The histograms were obtained using the LTFE density fields for \Stop{}(solid lines) and \Sbot{} (dotted lines). The blue lines indicate the void voxels, green for the wall-voxels, and red for the filaments and nodes. The \sinv{}-voxels are indicated with a purple dot-dashed line. }
    \label{fig:densitiespdf}
\end{figure}

For the resulting segmentation, in figure \ref{fig:densitiespdf} we show the probability distribution function (PDF) of the density field segregated according to cosmic environments. The histograms correspond to the LSDE densities estimated on a regular grid of spatial resolution $\Delta x \sim 0.48$\mpch. From \reffig{densitiespdf} we notice the quantitative effect of the smoothing scale: from \Stop{} (solid lines) to \Sbot{} (dotted lines), the    PDFs peak around larger values of density (by approximately a factor of  $0.5$ in $\log (\rho + 1)$ ). 
While the actual density values depend on the hierarchical scale, the qualitative conclusions to which we arrive are largely independent of the chosen scale and we see a distinction in the peaks of the distributions for each environment.  Due to the small number of nodes in our segmentation, we include those voxels together with filaments. The patterns that we appreciate follow the familiar conclusions shared by other cosmic web segmentation algorithms: voids are the less dense environments, followed by walls, and the denser environments correspond to filaments and nodes. However, it is also important to notice that we find important overlaps between the density PDF of different environments, which reinforces the known result that the density is not enough a criterion to select cosmic web environments. 
Equally important is to remark the fact that the PDF for the \sinv{} environment lies neatly in between the voids identified in \Stop{} (solid blue line) and the filaments+nodes (red dotted line) and walls from \Sbot{} (green dotted line).  
% ========================================================= 
% ==================  galaxies in voids  ==================
% =========================================================
%--------------FIGURE -----------------------
%------------------------------------------

\subsection{Galaxies tracing the cosmic web}

 %--------------------------------------------
%%%%%% =========================

%--------------------------------------------
%%%%%% =========================
\begin{figure}
\centering    \includegraphics[width=\columnwidth]{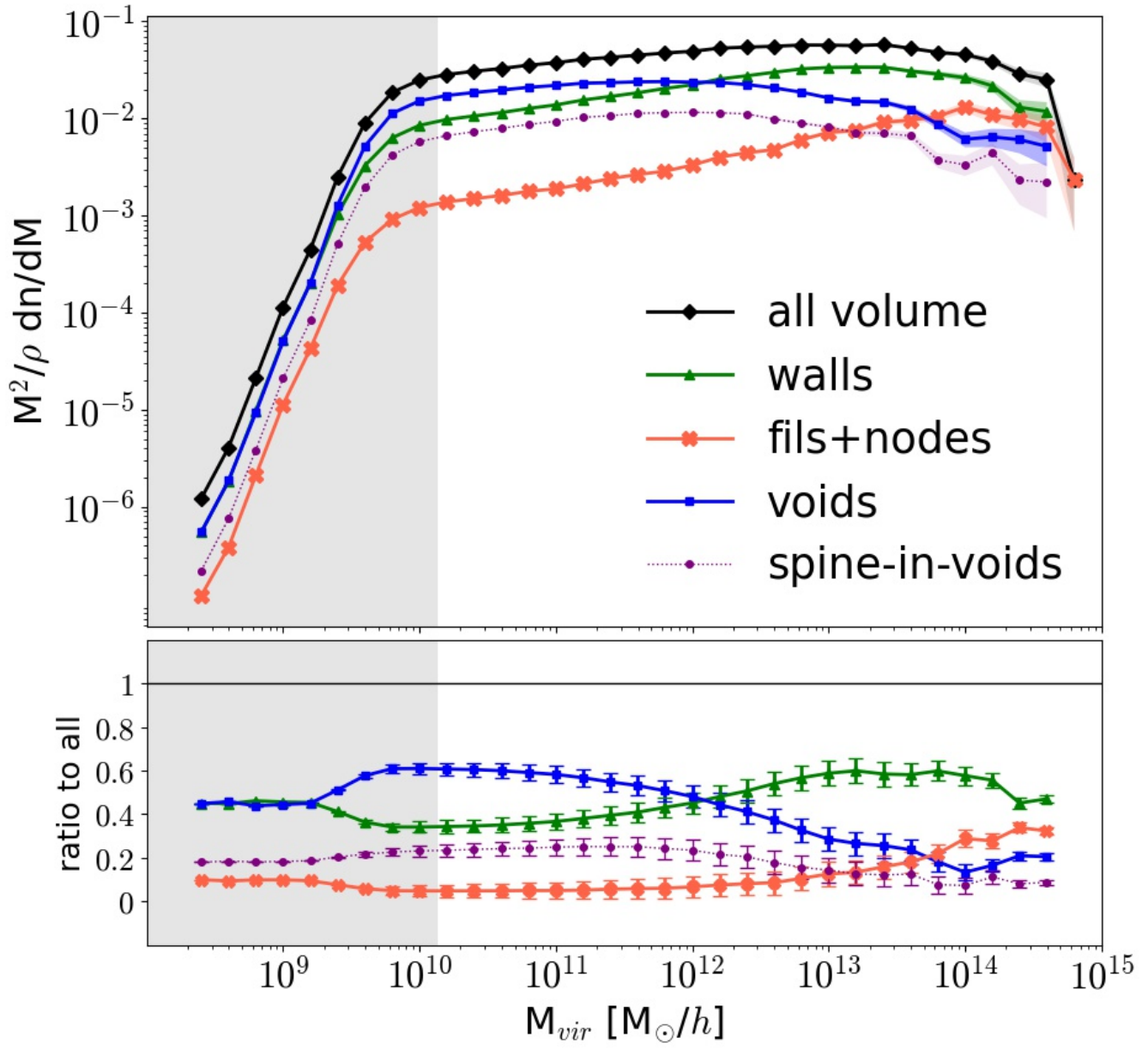}
    \caption{[Top] Differential halo number density as a function of mass for $z=0$. The co-moving number density of haloes less massive than $M$, $n(M, z=0)$,  is plotted here as the multiplicity function, $M^2\rho^{-1} dn/dM$. The solid black line is the result for the whole volume, the blue squares depict the haloes found in voids, the red crosses are for haloes found in nodes and filaments, the green triangles are for haloes found in walls, and the purple dots for haloes residing in the \sinv{}. The shaded regions indicate the Poisson errors for each halo mass function. The grey shaded region reflects the mass resolution for haloes of  $\geq 10^2$ particles. 
    [Bottom] The ratio of each environmental halo mass function with respect to the full sample, the error bars represent the relative Poisson errors for each mass bin.
     }
	\label{fig:hmf}
\end{figure}
%--------------------------------------------
%%%%%% =========================
%------------- table --------------------------
\begin{table} 
\caption{The volume filling fraction of voids, walls, filaments, and nodes, found in each environment for the three levels of the hierarchy, \Stop, \Smid, and \Sbot, represented by S1, S2, and S4, respectively.  The fractional number of galaxies per environment is indicated in ( ).
The fifth column indicates the number of galaxies found in the selected environment across all the hierarchy, this is, $\beta_{124} = \bigcup \beta_s $, the union of $\beta_s $ with $s = 1, 2, 4$ \mpch. The last column contains the fraction of galaxies found residing in the filamentary structure (nodes, walls, and filaments of \Stop) inside of voids at \Sbot\-\ or \sinv, indicated by $\sigma_{14}$. }\label{table:gals}
\centering
\begin{tabular}{|m{3.7em} m{1.1cm} m{1.1cm} m{1.1cm} m{1.1cm} m{1cm}|} 
 \hline
  & \texttt{S1} & \texttt{S2} & \texttt{S4}  & $ \beta_{124}$ & $\sigma_{14}$\\ [0.1ex]
 \hline\hline
 \textbf{Voids}  &  &  &  &  & \\[0.7ex]
   \small{Vol. frac.} & $77.03\%$  & $81.71\%$ & $86.97\%$ & --  & --\\[0.2ex]
  \small{Frac. gal.}  & ($44.5\%$) & ($48.7\%$) & ($58.6\%$) & ($31.9\%$) & --\\[0.1ex] \hline
     \textbf{Walls}  &  &  &  &  &\\[0.7ex]
  \small{Vol. frac.} & $20.31\%$ & $16.62\%$& $12.23\%$ & --  & --\\[0.2ex]
 \small{Frac. gal.}  & ($43.4\%$) & ($42.0\%$ )& ($35.9\%$ )& ($19.0\%$)& --\\[0.1ex]\hline
   \textbf{Filaments}  &  &  &  &  &\\[0.7ex]
  \small{Vol. frac.} & $2.19\%$& $1.37\%$ & $0.61\%$ & --  & -- \\[0.2ex]
  \small{Frac. gal.} & ($10.0\%$ )& ($7.7\%$) & ($4.5\%$) & ($0.8\%$) & --\\[0.1ex]\hline
   \textbf{Nodes}  &  &  &  &  &\\[0.7ex]
 \small{Vol. frac.}
 & $0.46\%$ & $0.31\%$ & $0.21\%$ & --  & -- \\[0.2ex]
  \small{Frac. gal.} & ($2.01\%$) & ($1.52\%$) & ($0.89\%$) & ($0.02\%$)& --\\[0.1ex]\hline
   \textbf{\sinv}  &  &  &  &  & \\[0.2ex]
\small{Frac. gal.} & - & - & - & - & \small{(22.8 $\%$)}\\\hline
\end{tabular}
\end{table}
%------------------------------------------

After running \sage\-\ with the parameters according to \cite{Croton16SAGE} for Bolshoi,  we identified a total of 7 909 721, or $\sim(199)^3$, galaxies at redshift $z=0$, fulfilling the condition of $M_*>0$. 
It is important to remember that the halo catalogues we used were created using the full-resolution Bolshoi simulation, and not the smoothed ones. 
From those,  approximately a third of the full sample (32$\%$), was found in voids across the three hierarchy levels,  $19\%$, were residing in walls, $0.8\%$ in filaments and $0.02\%$ in nodes. 
This corresponds to the fifth column of table \ref{table:gals}.  We have defined the full hierarchy, $ \beta_{124} = \bigcup \beta_s $,  with the condition that a given galaxy belongs to the same environment in all the three scales simultaneously. Roughly half of the sample matches this criterion. 

Now, to look at how these fractions distribute across the hierarchy, we also include the fractional number of galaxies per environment in \Stop, \Smid, \-\ and \Sbot.  See columns 2, 3, and 4 of table \ref{table:gals}, respectively. 
In this case the fractions add up to 100$\%$ in each level of the hierarchy.
From these numbers we notice the impact that the scale of description has in the classification of the  membership of galaxies.
For instance, at the top of the hierarchy we have almost the same percentage of galaxies located  voids ($44.5\%$) as we do find in walls ($43.4\%$).  With a yet smaller fraction of galaxies tracing filaments ($10\%$), and a minuscule population found in nodes, the densest environments ($2\%$). 
However, as we move through the hierarchy, from \Stop \-\ to \Smid \-\ and \Sbot, we can see how those fractions change. The fraction of galaxies found in voids increases (from $45\%$ to $49\%$ and finally, $59\%$), and, consequently, the fraction of galaxies in denser environments diminishes. 

This is a direct consequence of the change in the volume filling fractions of each environment. 
The volume filled by voids changes from $77\%$ to $87\%$, as we go from \Stop \-\ to \Sbot, and consequently, the volume filled by walls, filaments and nodes, decreases.

Now, in order to identify the population of galaxies tracing the inner structures of voids, we select the voxels identified as voids in the bottom of the hierarchy, retain their indices and then map those indices to the top of the hierarchy. Once this is done we can select the voxels that are simultaneously identified as ``non-void structures'' (filaments + walls + nodes) in \Stop{}, and as voids in \Sbot{}. 

We shall consider such defined population as tracing a distinct new environment,  which we dub \sinv,  and which comprises $23\%$ of the full galaxy sample (see last column of table \ref{table:gals}).

Having defined and described our galaxy samples split among our cosmic web environments, we now take a look on how different physical properties of galaxies vary across different population.
In what follows we consider \textbf{(A)} the halo-mass-function (HMF), \textbf{(B)} their spin, $\lambda$, \textbf{(C)} the stellar-to-halo mass (SHM) relation,  and \textbf{(D)}  their metallicities, in the context of the galaxies placement in the cosmic web. 

\subsubsection*{\textbf{A}. Environmental HMF}

The function $n(M, z)$ gives the comoving number density of haloes with a virial mass below $M_\mathrm{vir}$, denoted by $M$ to simplify the notation. In figure \ref{fig:hmf}, we show it as the halo multiplicity function  $M^2\rho^{-1} \mathrm{d}n/\mathrm{d}M$, with $\rho$, the mean density of the Universe, and also for each cosmic web type.  

The full sample is represented in black diamonds, while the ga\-laxies split by cosmic environment are represented as follows: red crosses for galaxies found in nodes and filaments, green triangles for walls, and blue squares for galaxies residing in voids. The purple dots represent the galaxies found in the \sinv \-\ environment.

Our halo mass functions cover 6 orders of magnitude in mass,  from $10^9-10^{15}[M_{\odot}h^{-1}]$. To ensure robustness in our analysis, we have set a minimum particle threshold for halo identification. 
We add a shaded region to indicate a halo mass convergence threshold, set at a minimum of 100 particles per group, which we believe is a conservative choice given that previous studies in the literature \citep{Behroozi13rockstar, Croton16SAGE} report a threshold of 20 and 50 particles per halo, respectively. 
However, the 100 particle threshold was found to be a reasonable value for the convergence of a simulated FoF groups HMF in \cite{warren_2006}. 
Therefore, we do not consider HMF, and so also the halo and galaxy populations, as converged below  masses of $1.35\times10^{10}$\Modoth.   

The first thing we notice from the environmental HMF split is that the galaxies at the low mass end are found mostly residing in voids; this is, for  masses up to $10^{12}$\Modoth, at which point, the number of galaxies residing on walls start to dominate the high mass end. 

Another feature to note is the low number counts of galaxies residing in the densest environments such as filaments and nodes. We attribute this results as due to a specific systematic induced by the classification scheme. As detailed in subsection \ref{subsec:cweb}, the voxels residing in the border between voids were marked with a morphological label, which, depending on the number of different neighbouring voids, were classified as a walls ($\mathcal{N}_v=2$), filaments ($\mathcal{N}_v=3$) or nodes ($\mathcal{N}_v\geq 3$). In each case, the environments are 1-voxel thick, resulting in vanishingly small volume-filling fractions leading to  artificially lower number of galaxies identified in such environments. This is clearly demonstrated by the volume-filling fractions quoted in table \ref{table:gals}, where we can see that the volume occupied by filaments fluctuates between 2.19$\%$ and 0.61$\%$, while the volume occupied by nodes is practically negligible. For this reason we plot the halo mass function for filaments and nodes jointly. 

However, a difference between environments can be noticed through the  increasing steepness of the HMF for the densest environments (filaments+nodes) at the high mass end, more pronounced than the one for walls, and in opposite direction than the HMFs for haloes residing in voids and in the \sinv. 
Respect to the latter,   their HMF follows the same behaviour as the haloes found in voids but with a lower number count. As indicated in the last column of table \ref{table:gals}, this sample comprises 23$\%$ of the full sample.  

The lower-panel of figure \ref{fig:hmf} illustrates the relative fraction of haloes found in each one of the environments with respect to the full volume.  
A striking feature is the threshold value for the mass at around $10^{12}$\Modoth, at which the fraction of haloes residing in walls starts to dominate at the high-mass end.  
For haloes with masses below this value, the dominant population is residing in voids.  
The fraction of haloes found in the densest environments is heavily suppressed during the full mass range, except by the last mass bin, at $M_\mathrm{vir}\sim10^{15}$\Modoth, where the full sample resides in nodes. For a cleaner visualisation we omit the last bin in the ratios depicted in the bottom panel of figure \ref{fig:hmf}. 

\subsubsection*{\textbf{B.} Spin}

\begin{figure}
    \centering
    \includegraphics[width=\columnwidth]{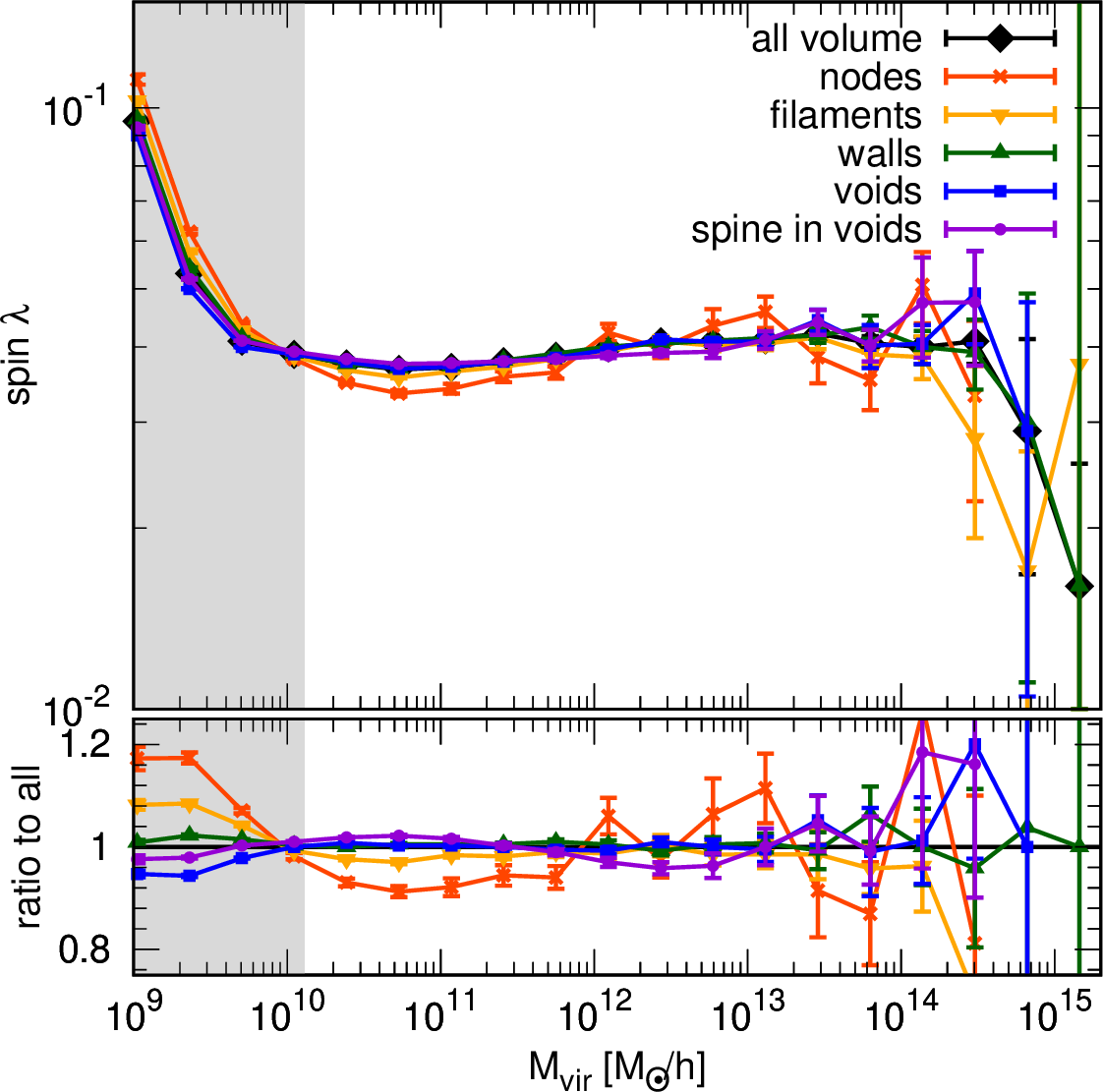}
        \label{fig:spin-cweb}
    \caption{[Top] Value of the spin parameter, $\lambda$, for the galaxies located in different cosmic environments. In black diamonds we indicate the full sample, red crosses, haloes found in nodes, green triangles are for walls, inverted yellow triangles are filaments, blue squares represent voids and the purple dots, the haloes found in the \sinv.  %
   [Bottom] Ratio of each environment with respect to the full sample. The shaded gray areas represent the mass resolution limitation of 100 particles per halo. The errors indicate a bootstrap on the medians per mass bin. 
}
    \label{fig:spin}
\end{figure}

Another relevant property for dark matter haloes, besides their mass is their internal structure. 
One of the physical properties describing the internal structure of dark matter haloes is the spin parameter, which provides a measure of the net internal rotation of the dark matter halo and is typically provided in terms of their angular momentum.  
\sage\-\ follows the definition from \cite{Bullock_spin} for the spin parameter of dark matter halos, $\lambda=\lvert J\rvert/\sqrt{2}R_\mathrm{vir}V_\mathrm{vir}$, where $|J| = S_x^2+S_y^2+S_z^2$ is the halo total angular momentum, and  $R_\mathrm{vir}$ and $V_\mathrm{vir}$, the virial radius and virial velocity of the halo, respectively. 

In figure \ref{fig:spin} we show the spin parameter of the fraction of haloes found in each of the LSS environments in a given mass bin. The bottom panel shows the relative difference between each environment and the total volume.  In this case the galaxies split by cosmic environment are: red crosses for the node-halo sample, green triangles for walls, yellow inverted triangles for filaments,  blue squares for void galaxies, and  purple dots haloes found in the \sinv \-\ environment.

For each halo mass bin, we estimate the uncertainties through a bootstrap of the median, using 100 bootstrap samples. 
We found that the overall shape of the spin as a function of $M_\mathrm{vir}$ is relatively flat around the value  $\lambda=0.04$. However, when we analyzed the relative ratio of spin properties for each cosmic web halo population with respect to the all-volume sample, we observed some environmental effects,  for haloes with masses between $10^{10}-10^{12}$\Modoth, and peaking around $10.4-10.6\times10^{10}$\Modoth, the range for which we have the strongest environmental effects.

In this mass range there is a clear trend showing that haloes found in the densest environments get a suppression of their spin, while the haloes in the less dense environments such as voids or \sinv\-\, have an enhancement in their angular momentum.  In particular, we found that the spin of haloes in nodes, which are the densest regions of the cosmic web, exhibits a suppression of approximately 10$\%$ compared to the full sample. Similarly, we observed a suppression of spin for haloes in filaments, although of smaller proportions.  The haloes in walls seem to be unaffected with respect to the overall sample, but those found in the less dense environments, do show an increase of their spin. This effect was recently observed around Milky Way analogues located in walls \citep{Aragon23-MW}. 
Interestingly, this enhancement is stronger for the haloes in the \sinv\-\ environment than for the haloes in voids.

For haloes with  $M_\mathrm{vir}>10^{12}$\Modoth, we did not observe any significant patterns in the spin properties with respect to the cosmic web environment. 

\subsubsection*{\textbf{C.} Stellar-to-halo mass relation}

\begin{figure}
    \centering
    \includegraphics[width=\columnwidth]{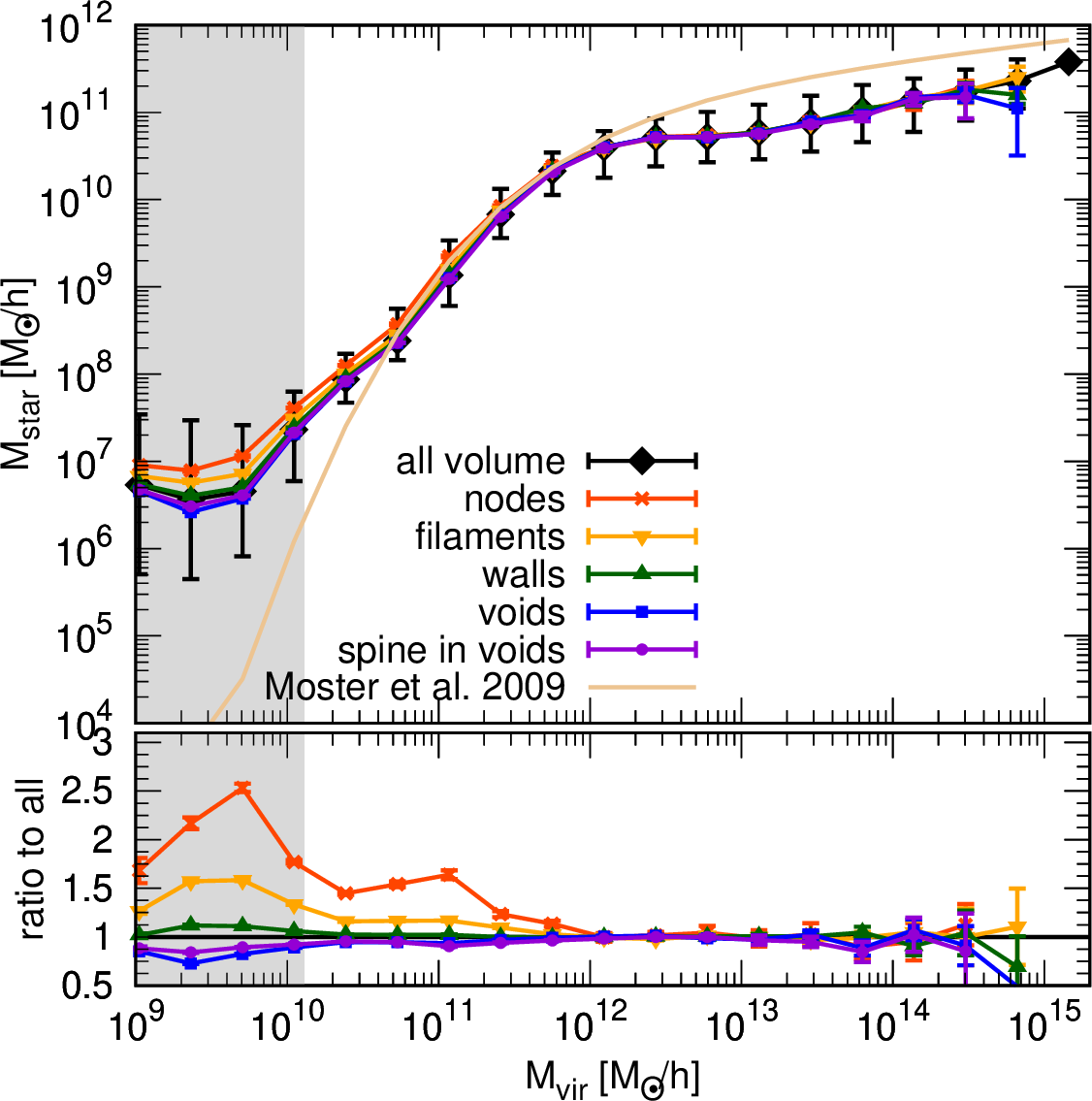}
        \caption{[Top] Stellar-to-halo mass relation, $M_{\star}$ - $M_\mathrm{vir}$ as function of the environment. Black diamonds
        indicate the full sample, red crosses, haloes found in nodes, green triangles are for walls, inverted yellow triangles
        are filaments, blue squares represent voids and the purple dots, the haloes found in the \sinv. 
        The solid golden line reflects the prediction of \citet{Moster_2010}.
        [Bottom] Ratio of each environmental SHM function with respect to the full sample. The shaded gray areas represent 
        the mass resolution limitation of 100 particles per halo. The errors indicate a bootstrap on the medians 
        per mass bin.
        }
    \label{fig:mstar}
        
\end{figure}

The link between galaxy properties and their host DM haloes can be understood via the stellar-to-halo mass (SHM) relation.  
To demonstrate the effect of the environment on the SHM function, we plot the stellar mass, $M_*$, as function of the mass of the host halo, $M_\mathrm{vir}$, in figure \ref{fig:mstar}, for the full sample, and also for each cosmic environment with the same legends as explained above. 

We compare our results with the parametric fit proposed by \cite{Moster_2010}, whose estimation for the stellar mass agrees with the trend of our sample in the halo mass range $10.3\times10^{10}-10.1\times10^{12}$ [\Modoth]. 

To emphasize the effect of the environment on the SHM relation, in the bottom panel of \ref{fig:mstar}, we plot the ratio of each individual cosmic environment to the total galaxies sample.

A very interesting feature arises from analyzing these ratios. A clear dependence on the environment is present for haloes below the mass threshold of $M_\mathrm{vir}<10^{12}$\Modoth.  
The  environmental dependence on the SHMR is the most prominent for the node-galaxy population, in which case we see an enhancement above  $50\%$ with respect to the stellar mass of the full sample. 
The galaxies in filaments, we can also observe an increase in their stellar masses, close to $15\%$.  The maximum departure is found for galaxies in haloes of masses around $10^{11}$\Modoth.
As for the galaxies found in walls, we find no significant departure from the overall SHMR, and a small decrease of stellar masses is seen for the galaxy samples residing in voids and in the \sinv\-\ environment. 

For galaxies hosted in haloes of masses above the threshold, $M_\mathrm{vir}>10^{12}$\Modoth,  the environmental split seems to be a subdominant effect on their stellar content.

%%%%%%%%%% metals %%%%%%%%%

\subsubsection*{\textbf{D.} Metallicity}

\begin{figure}
    \centering
      \begin{subfigure}[b]{0.4\textwidth}
        \includegraphics[width=\textwidth]{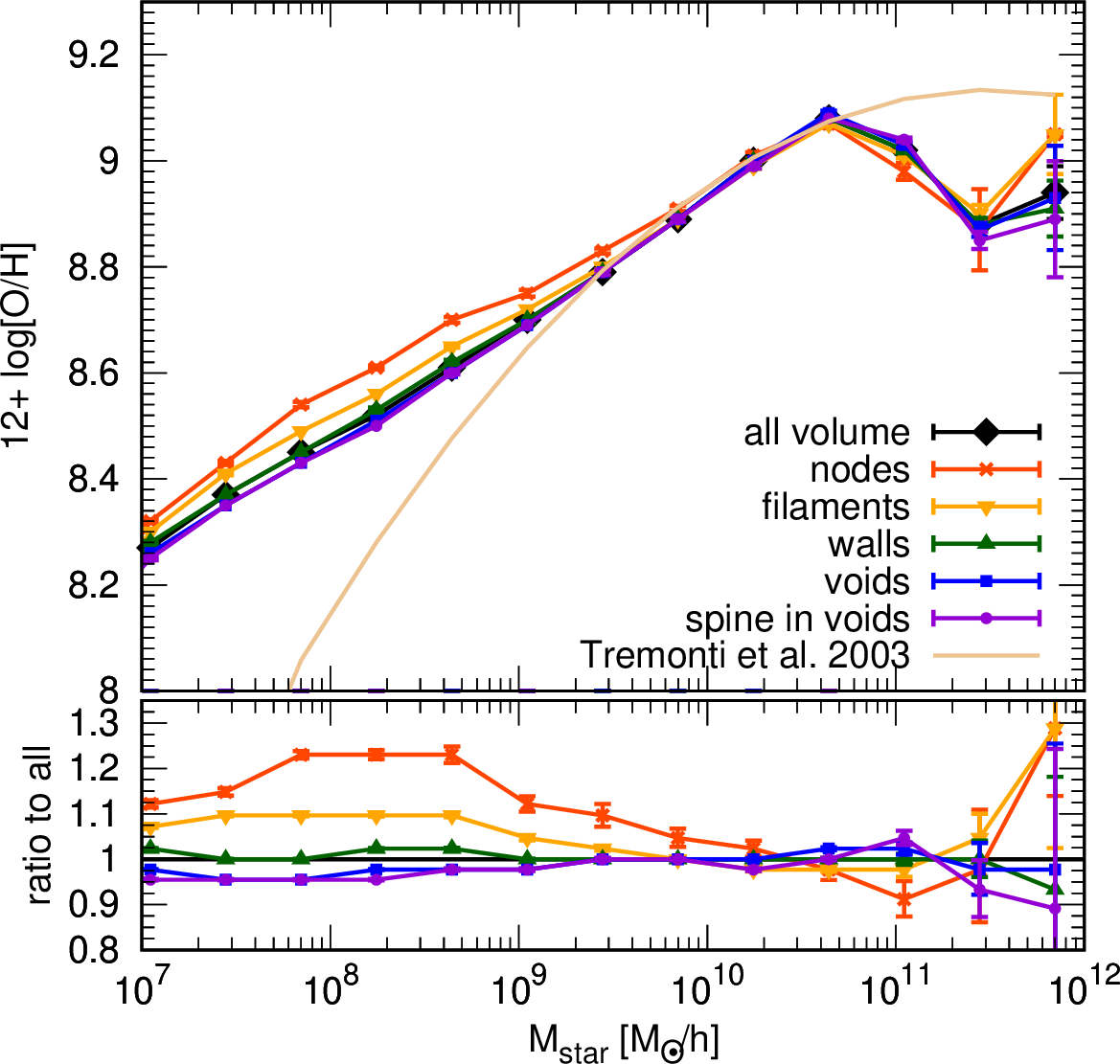}
    \end{subfigure}
   \caption{[Top] Metallicity as function of stellar mass, $M_{*}$. We report the ratio of Oxygen to Hydrogen in units of $12 + \log(O/H)$ and compare our results with the formula presented in \citet{Tremonti_2004} (golden line). In black diamonds we indicate the metallicities of galaxies for full sample, red crosses are for haloes found in nodes, green triangles for walls, inverted yellow triangles are filaments, blue squares represent voids and the purple dots are for the haloes found in the \sinv. [Bottom] The relative difference per each environment with respect to the full sample.}
    \label{fig:metal}
\end{figure}

Another property of galaxies helping bridge our understanding between galaxy formation and evolution with their cosmic web location, is the metal contents of galaxies.
The metal content or metallicity, expressed as $12+\log_{10}(O/H)$, provides a measure of the efficiency of stars in forming metals. 
Figure \ref{fig:metal} shows the metal fraction as function of stellar mass, split by environment. 

The main trend of all of our samples corresponds to an increasing function of metal content for higher stellar masses, which lies in agreement with the observational result reported by \cite{Tremonti_2004}, for galaxies with stellar masses in the range of $1.1\times10^9-1.2\times10^{10}$\Modoth. 

Beyond this overall trend, the environmental dependence is noticeable. 

The densest environments, corresponding to nodes and filaments of the cosmic web, exhibit a pronounced enhancement of their metallicities compared to the full sample. 
This enhancement reaches a maximum of ~25$\%$, for stellar masses of $10^8-10^9$\Modoth, declining and vanishing for $M_*\sim10.2^{10}$\Modoth. The filaments show a very similar yet less prominent effect, while the metallicities of galaxies residing in walls show no difference with respect to the overall sample. As for the low density environments, such as voids and \sinv, we see a reduction of their metallicities of almost 10$\%$, for masses in the same range, $M_*<10^{10}$\Modoth. 
In this case, we see that the population of galaxies residing in the \sinv\-\ environment suffer from the same reduction of their metal fraction in comparison with the full sample, as do our void-galaxies sample. 
Galaxies with higher stellar masses than $10^{10}$\Modoth, do not have a discernible trend in their metallicities with respect of the cosmic web environment where they reside.

As we could see in our previous result, the stellar mass of galaxies shows a discernible dependence on the cosmic web environment, in particular, a trend of increased stellar mass for the galaxies residing in the denser environments, and a reduced gas content for galaxies in voids.
It is well known that the relationship between stellar mass and metallicity works as a diagnostic of the physical processes that govern galaxy formation and evolution. In \sage, this connection is modelled assuming a yield, $Y$, or fraction  of the stellar mass that is converted into metals during the process of stellar evolution.  
Hence,  the environmental dependence of the SHMR, gets imprinted in the metallicities-$M_*$ plane with similarly distinctive signatures.
  
% %%%%%%%%%% metallicity %%%%%%%%%

% ========================================================= 
% ====================   DISCUSSION   ====================
% =========================================================
\section{Discussion}
\label{sec:discussion}

In this paper we have presented an in-depth analysis  of the connection between galaxy properties and the internal structure of voids using cosmological simulations.
While some works have extensively looked at both factors separately, we aimed  to study the interplay between both. 

The ability to select and describe the inner structure of voids relied on the implementation of appropriate filters selecting specific scales ($\beta_s = 1, 2, 4$\mpch) in the gravity-only simulations and constructing a hierarchy of voids, walls, filaments and nodes, similarly to the work of \cite{Aragon_Calvo_2012}.

However, by further populating the DM haloes with galaxies, using the numerical implementation of the \sage\-\ method, we were able to pose the question of what population of galaxies traces the internal filamentary structure of voids, or \sinv, as we named such environment, in terms of their halo mass function, spin, stellar mass, and metallicity.  
While hydrodynamical simulations offer a more detailed treatment of baryonic processes and their coupling with larger-scale structures, it is vital to note that their use can be computationally expensive and may introduce their own set of biases. Our approach, utilizing dark matter-only simulations and semi-analytical modeling, is a well-established and widely used methodology in computational cosmology, as tested in the rich literature \citep[see e.g.][]{2002MNRAS.337.1193M, 2003MNRAS.340..160B, 2008MNRAS.387..128P, 2009MNRAS.396.1815F,2012MNRAS.425.2049H,2017MNRAS.469..594B,2020A&A...638A..60A,hellwing21,2021MNRAS.503.2280G}.

To characterize such hierarchical structure of voids, we presented the void size functions and density profiles, as function of the smoothing scale, $\beta_s$.  Our results agree with previous results regarding the density profile of voids identified using watershed techniques \citep[see for instance][]{van_de_Weygaert_2014, shed_voids1}, but they also show the effect of the smoothing scale, $\beta_s = 1, 2, 4$\mpch.  

With respect to the void size functions, we found that the peak of the PDFs scales with the smoothing scale through a simple relationship as: $R_\mathrm{peak}\sim 11\beta_s^{0.6}$\mpch.  

We found a strong environmental dependence on our HMFs (see figure \ref{fig:hmf}). A threshold of \MthUnit marks the turnover between haloes in voids and haloes in walls dominating the number counts, and it is present prominently in the HMFs. 
We adjudicate the low number counts of haloes in filaments and nodes to the morphological labeling in our cosmic web segmentation, in particular to artificially thin (one-voxel thick) environments other than voids. 
Conversely, our number of void-galaxies is quite large (45-59$\%$), which we attribute to the watershed technique being a volume filling algorithm, which by construction returns too thin filaments, and potentially leads to mislabelling galaxies at the edge of the voids as void-galaxies instead of wall, filaments or node-galaxies. 

The \sinv{} environment shows the same behaviour as the void-galaxy sample, but with suppressed number counts, signaling that the dominating effect is the one of being found in a low-density environment, just as void-galaxies are. 

Interestingly enough, the other quantity related to describe the properties of dark matter haloes, \emph{i.e.} the spin parameter $\lambda$, shows environmental effects for a mass below the same threshold \MthUnit. In this case, however, we find that the haloes located in the densest regions of the cosmic web have a smaller spin parameter compared to those in voids.  This suppression is stronger for haloes in nodes and to a lesser extend for those in filaments, suggesting that the tidal forces exerted by the surrounding large-scale structures play a role in shaping the spin of haloes.  
The haloes found in the under-dense regions, on the other hand, show a slight enhancement in their spin parameter with respect to the trend of the full sample.  
Our results lie in agreement with the work of \cite{elaphro}, where similar void-finding  and halo identification techniques were used. We found the overall shape of the  $\lambda(M_\mathrm{vir})$ function to be relatively flat around the value  $\lambda=0.04$, which is a slightly higher value than the one reported as the peaks of the distributions reported by \cite{vladimir}. 
Nevertheless we agree in reporting higher values of $\lambda$ for void galaxies, and smaller for galaxies in high density environments. 

However, when holding our results against those reported by \cite{hellwing21}, the comparison is not as direct, as we have probed different mass ranges in our halo catalogues. 
While they report a similar decrease in the spin parameter for their node-haloes, in the mass range of  $M_\mathrm{vir}\sim10.6\times10^{7}-10^{10}$\Modoth, and also a minimal enhancement for the spin of their void-haloes, in the same mass range, they have a substantial further decrease of the spin of the void sample. This reduction of spin is of almost 35$\%$, as they quote, and is present for haloes of masses around $10^{11}$\Modoth in $M_\mathrm{vir}$.  
Care should be taken when attempting a direct comparison, though, as we have different mass ranges of our halo samples. For instance, we do have void-haloes at all mass ranges, whereas, the authors of this work only have them for masses below $10.4\times10^{11}$\Modoth. As our cosmic web segmentation methods are different (the authors have implemented NEXUS which is a Hessian-based method, that classifies the cosmic web based on the tidal tensor signatures), and also our halo identification schemes differ, discrepancies are to be expected. 
However, the main trend hinting to a suppression of $\lambda$ for haloes residing in the densest environments is present in both our works, as well as the detection of a mass threshold at which the various cosmic web populations deviate from the mean trend describing the overall halo population.  In our work, this mass threshold is of the order $M_\mathrm{th}\sim 10^{12}$\Modoth, over two orders of magnitude higher than the mass-scale reported by \cite{hellwing21}.
An in depth analysis of the spin alignment with respect to the cosmic web has been carried out in a series of papers \citep{2018MNRAS.481..414G,2019MNRAS.487.1607G,2021MNRAS.503.2280G} also using NEXUS and variants of the method. Such studies confirmed a mass-dependent spin–filament alignment, originally proposed by \citep{2007ApJ...655L...5A} and observationally  studied by e.g. \cite{2013MNRAS.428.1827T}.

Understanding the interplay between the cosmic web and the properties of galaxies is crucial in modern cosmology. While dark matter haloes provide a convenient framework to study the formation and evolution of galaxies, they are not observable, and therefore it is important to go beyond their properties and analyze the stellar content and metallicities of galaxies in terms of their placement on the cosmic web. 

In this case, our major finding shows that our stellar-to-halo mass functions reflect a strong environmental dependence for masses below \MthUnit, the same mass threshold as the one reported for halo properties. In particular, we see a significant increase of s\-te\-llar masses for haloes in nodes and filaments, in the range  $M_\mathrm{vir}<M_\mathrm{th}$.
For the galaxies identified in the under-dense regions, we have a decreased stellar mass ratio in the same mass range, in agreement with previous works (see for instance \cite{HoyleLF2005, croton_2005, Moorman2015}).  In contrast, 
\cite{2016_Beygu,2022_DominguezGomez} have reported no differences in observed stellar masses between void and non-void galaxies, while the numerical results from \cite{2020A&A...638A..60A, 2020MNRAS.493..899H, 2022MNRAS.517..712R, Gal_rraga_Espinosa_2023} and observational analysis of  \cite{2021_Florez, 2023Natur.619..269D} also agree with our findings of decreased stellar masses of void galaxies.  

As for the metallicities or the abundance of elements heavier than helium, usually expressed as the ratio of the mass of these elements to the mass of hydrogen, the major trends follow the same pattern: a strong environmental dependence showing that galaxies in high density environments (nodes, filaments, and walls) have sharp increases in their metal abundances, with respect to the full sample, while the galaxies populating the low-density environments (voids and \sinv) presented a decreased metallicity value as function of $M_*$. In this case we observe that the environmental effects are present for masses below $M_* \sim10.1\times10^{10}$\Modoth. 
Interestingly, we find that the population \sinv\-\  shows a  slightly stronger suppression of metallicities, when compared to the full sample.

The previous results are consistent with the expectation that galaxy mergers and interactions are more frequent in denser environments, leading to a higher star formation rate and, thus, greater rate of metal production. 
This expectation aligns with the idea of nodes and filaments having a higher probability of hosting massive star-forming galaxies and clusters, which are known to be efficient in producing metals through their stellar winds and supernova explosions.  
Our findings provide further evidence that the cosmic web plays a critical role in the formation and evolution of galaxies, with the denser regions being more conducive to metal formation.

Similar to the works concerning the environmental dependence of the stellar mass, different authors report contrasting results in terms of the metal content of void galaxies. For instance \cite{Kreckel_2014} report no significant differences between field and void galaxies and a recent study using the EAGLE simulation conducted by \cite{2022MNRAS.517..712R} found higher metallicities for massive galaxies. The latter team found their results consistent with the fact that inner void galaxies are more isolated than those in denser environments.  
The origins of these discrepancies can be due to  a number of factors, from differences in void finding techniques \citep[see the discussion in][]{aspen_amsterdam, libeskind_comparison, shed_voids1}, to the statistical errors associated to different samples, and even more, the different biases linked to the tracers used by each team (e.g. \cite{2015MNRAS.454..889N, 2015MNRAS.454.2228N, 2017MNRAS.470.4434P,2021_Florez}).

Furthermore, even if metallicity is not a perfect indicator of a galaxy's age (for instance, a galaxy can undergo episodes of enhanced star formation that can boost their metallicity without significantly increasing its age, but also periods of gas accretion which can  dilute their metallicities without affecting the age), it is well known that it can be used as a proxy for a galaxy's age as it is easier to determine the abundance of heavy elements observationally, as to directly measure the age. 
From this it follows that our results support the picture of void galaxies being younger (bluer), and galaxies in nodes or filaments, older (redder). 
Respect to the \sinv-galaxy sample, we conclude that their properties follow those of the void-galaxies in general. In the case of their metallicities, we found a slight decrease in their metal fraction with respect to the full sample.
We interpret this further suppression as an effect of the more frequent gas accretion periods, which can be fed via the filaments connecting such galaxies with the boundaries of the voids, and diluting their metallicities.

% ========================================================= 
% ====================   CONCLUSIONS   ====================
% =========================================================

\section{Conclusions}
\label{sec:conclusions}

We analyzed the link between a galaxy's location in the cosmic web and its physical properties such as their morphology (spin), size (mass), stellar mass ($M_*$), and metallicity. Our methodology allowed us to select the filamentary structure in the interior of cosmic voids and to characterize the  sample of galaxies tracing such cosmic web environment, which we named \sinv. 

The cosmic web segmentation was done by implementing the watershed algorithm, or \spine on a set of gravity-only simulations which were a lower resolution of the Bolshoi cosmological simulation. The galaxies were added by implementing the semi-analytical galaxy evolution code, \sage, using the available halo-trees for the full-resolution Bolshoi simulation. Under this scheme, the merger trees are extracted from a DM-only $N$-body simulation  and gas
processes are treated with semi-analytic recipes.

The specific segmentation, even if parameter-free, introduced a systematic effect in artificially low number counts for the densest environments, particularly visible in our HMFs.
Nevertheless, our results show that the stellar mass and metallicity of galaxies are significantly influenced by their location within the cosmic web, as well as the halo properties such as their mass functions and spin.  

An interesting result we found throughout our analysis was a threshold mass: \MthUnit, fixing the mass range for which we see environmental dependence in the galaxies properties. This value is in agreement with the expectation of haloes with masses larger than $10^{12}$\Modoth, to be predominantly in dense environments.
In particular, for the halo mass functions, this $M_\mathrm{th}$ signals the turnover between void-haloes and wall-haloes dominating the number counts in the HMFs. 
For the spin parameter, $\lambda$, we see an environmental dependence for haloes below this $M_\mathrm{th}$. Particularly, our results show that haloes in denser environments get a suppression of their spin with respect to the full sample, whereas haloes in under-dense regions suffer a slight increase in their angular momentum. We interpret this result as the effect of  the tidal forces exerted by the surrounding large-scale structures, playing a role in shaping the spin of haloes.  
This same threshold gets imprinted in the stellar content of the galaxies.

An interesting follow up would involve using the \spine{} framework to investigate how galaxy properties vary based on their proximity to distinct locations within the cosmic web. This approach aligns with previous studies like those conducted by e.g. \citep{2015MNRAS.452.3369C, 2018MNRAS.474..547K}. 
For example, the authors of \cite{elaphro} have searched into the different definitions of void center to quantify the effect on the halo properties, using the gravitational potential to identify the center of their voids and more recently, the work of \cite{2023MNRAS.525...91P}, proposes a new void identification by looking at geometric-optics parameters for gravitational lensing analysis.
Others, for instance \cite{Tavasoli2021}, have quantified the effect in the geometrical position of void galaxies, looking at density profiles of stacked voids which can be extended  using our segmentation method to further to constrain the values of cosmological parameters as in \cite{ap_voids, ap_voids_sdss}. Also, the authors of \cite{Donnan_2022} report finding a trend between the metallicity of galaxies as function of their distances to filaments detected in simulations. As  all of the aforementioned works use different methods and data, and we know that there are a diversity of results regarding the connection between galaxy formation an evolution versus their cosmic web placement, we think that this as future analysis would be a worthwhile path of research. 

To summarize, we have shown that the properties of galaxies tra\-cing the internal structure of voids can be described to have smaller stellar masses and to be located in haloes with smaller $M_\mathrm{vir}$, to have an enhanced spin, and a lower metallicity. 
Our findings provide further evidence that the environmental effect is present for certain threshold in mass. Specifically, we found that the cosmic web plays a critical role for galaxies with $M_{\mathrm{vir}}\sim10^{11}$\Modoth and $M_*\sim10^8$\Modoth.
These results altogether point to the need of testing our framework on higher resolution simulations to explore the effect of cosmic web environment on satellite galaxies.

\section*{Data Availability}
The halo trees used in this work are publicly available at this \href{https://www.slac.stanford.edu/~behroozi/Bolshoi_Trees/}{URL}, and the numerical implementation of the density estimator can be found in \href{https://github.com/miguel-aragon/Lagrangian-Sheet-Density-Estimator}{this repository}. The version of the code \sage\-\ can be downloaded from \href{https://github.com/darrencroton/sage}{\sage.} The \spine code and  data underlying this article will be shared upon reasonable request to the corresponding author. 

\section*{Acknowledgements}
The authors thank Anatoly Klypin for providing the numerical re-sampled Bolshoi simulations used for our hierarchical analysis.
The authors thank Boudewijn F. Roukema for helpful suggestions that improved the analysis presented in this work, and Darren Croton for his guidance regarding the \sage\-\ output.
We also wish to thank the anonymous referee and editor for their careful read of our manuscript and helpful suggestions.
MJ acknowledges the support of the Polish Ministry of Science and Higher Education MNiSW grant DIR/WK/2018/12, as well as the research project “VErTIGO” funded by the National Science Center, Poland, under agreement number 2018/30/E/ST9/00698.
M.A acknowledges support for the DeepVoid project and this publication from grant 62177 from the John Templeton Foundation.
WAH acknowledges the support from the research project "COLAB" funded by the National Science Center, Poland, under agreement number UMO-2020/39/B/ST9/03494.
MP acknowledges that part of this work was supported by Universitas Copernicana Thoruniensis in Futuro under NCBR grant POWR.03.05.00-00-Z302/17 and by the NCN grant 2022/45/N/ST9/01698.
Part of this work was supported by the ``A next-generation worldwide quantum sensor network with optical atomic clocks'' project, which is carried out within the TEAM IV program of the Foundation for Polish Science co-financed by the European Union under the European Regional Development Fund. 
OV thanks the support from UNAM DGAPA/PAPIIT Grants IG101222 and IG102123.
We also acknowledge the help provided by J.C. Clemente Gonz\'alez to make use of the computing facilities of the LAMOD UNAM project through the cluster Atocatl (\href{http://www.lamod.unam.mx/}{http://www.lamod.unam.mx/}). LAMOD is a collaborative effort between the IA, ICN, and IQ institutes at UNAM. We also thank Roman Feiler to handle the trees data via the computational resources of the Astronomy Institute of NCU.

\bibliographystyle{mnras}
\bibliography{spine}

% Don't change these lines
\bsp	% typesetting comment
\label{lastpage}
\end{document}